%% file: hcomp-bestarms-sideinfo.tex
\documentclass{article}
\input{commands}
\title{Learning to Hire Teams\thanks{Short version of this paper will appear in HCOMP'15.}\\}
\author{
Adish Singla\\
ETH Zurich\\
adish.singla@inf.ethz.ch \\
\And
Eric Horvitz\\
Microsoft Research\\
horvitz@microsoft.com \\
\And
Pushmeet Kohli\\
Microsoft Research\\
pkohli@microsoft.com \\
\And
Andreas Krause\\
ETH Zurich\\
krausea@ethz.ch \\
}
\begin{document}
\maketitle
%%%%%%%%%%%%%%%%%%%%%%%%%%%%%%%%%%%%%%%%%%%%%%%%%%%%%%%%%
%%%%%%%%%%%%%%%%%%%%%%%%%%%%%%%%%%%%%%%%%%%%%%%%%%%%%%%%%
\begin{abstract}
\begin{quote}
Crowdsourcing and human computation has been employed in increasingly sophisticated projects that require the solution of a heterogeneous set of tasks. We explore the challenge of building or hiring an effective team, for performing tasks required for such projects on an ongoing basis, from an available pool of applicants or workers who have bid for the tasks. The recruiter needs to learn workers' skills and expertise by performing online tests and interviews, and would like to minimize the amount of budget or time spent in this process before committing to hiring the team. How can one optimally spend budget to learn the expertise of workers as part of recruiting a team? How can one exploit the similarities among tasks as well as underlying social ties or commonalities among the workers for faster learning?  We tackle these decision-theoretic challenges by casting them as an instance of online learning for best action selection. We present algorithms with PAC bounds on the required budget to hire a near-optimal team with high confidence. Furthermore, we consider an embedding of the tasks and workers in an underlying graph that may arise from task similarities or social ties, and that can provide additional side-observations for faster learning. We then quantify the improvement in the bounds that we can achieve depending on the characteristic properties of this graph structure.  We evaluate our methodology on simulated problem instances as well as on real-world crowdsourcing data collected from the oDesk platform. Our methodology and results present an interesting direction of research to tackle the challenges faced by a recruiter for contract-based crowdsourcing.
\end{quote}
\end{abstract}
%\category{H.1.2}{Information Systems}{User/Machine Systems}
%\terms{Algorithms, Economics, Experimentation}
%\keywords{Crowdsourcing, budgeted exploration, multi-armed bandits, value of information, procurement auctions, PAC bounds}
%%%%%%%%%%%%%%%%%%%%%%%%%%%%%%%%%%%%%%%%%%%%%%%%%%%%%%%%%
%%%%%%%%%%%%%%%%%%%%%%%%%%%%%%%%%%%%%%%%%%%%%%%%%%%%%%%%%
\input{introduction}

%%%%%%%%%%%%%%%%%%%%%%%%%%%%%%%%%%%%%%%%%%%%%%%%%%%%%%%%%
%%%%%%%%%%%%%%%%%%%%%%%%%%%%%%%%%%%%%%%%%%%%%%%%%%%%%%%%%
\input{related}
\input{model}

%%%%%%%%%%%%%%%%%%%%%%%%%%%%%%%%%%%%%%%%%%%%%%%%%%%%%%%%%
%%%%%%%%%%%%%%%%%%%%%%%%%%%%%%%%%%%%%%%%%%%%%%%%%%%%%%%%%
\input{algorithm}

%%%%%%%%%%%%%%%%%%%%%%%%%%%%%%%%%%%%%%%%%%%%%%%%%%%%%%%%%
%%%%%%%%%%%%%%%%%%%%%%%%%%%%%%%%%%%%%%%%%%%%%%%%%%%%%%%%%
\input{analysis}

%%%%%%%%%%%%%%%%%%%%%%%%%%%%%%%%%%%%%%%%%%%%%%%%%%%%%%%%%
%%%%%%%%%%%%%%%%%%%%%%%%%%%%%%%%%%%%%%%%%%%%%%%%%%%%%%%%%
\input{experiments}

%%%%%%%%%%%%%%%%%%%%%%%%%%%%%%%%%%%%%%%%%%%%%%%%%%%%%%%%%
%%%%%%%%%%%%%%%%%%%%%%%%%%%%%%%%%%%%%%%%%%%%%%%%%%%%%%%%%
\input{conclusion}

%%%%%%%%%%%%%%%%%%%%%%%%%%%%%%%%%%%%%%%%%%%%%%%%%%%%%%%%%
%%%%%%%%%%%%%%%%%%%%%%%%%%%%%%%%%%%%%%%%%%%%%%%%%%%%%%%%%
\clearpage
{
%\fontsize{9.5pt}{10.5pt}
%\selectfont
\bibliographystyle{aaai}
\bibliography{hcomp-bestarms-sideinfo}  % sigproc.bib is the name of the Bibliography in this case
}
%%%%%%%%%%%%%%%%%%%%%%%%%%%%%%%%%%%%%%%%%%%%%%%%%%%%%%%%%
%%%%%%%%%%%%%%%%%%%%%%%%%%%%%%%%%%%%%%%%%%%%%%%%%%%%%%%%%
%%%%%%%%%%%%%%%%%%%%%%%%%%%%%%%%%%%%%%%%%%%%%%%%%%%%%%%%%%%%%%%%%%%%%%%%%%
%\clearpage
%\input{appendix}
%%%%%%%%%%%%%%%%%%%%%%%%%%%%%%%%%%%%%%%%%%%%%%%%%%%%%%%%%
%%%%%%%%%%%%%%%%%%%%%%%%%%%%%%%%%%%%%%%%%%%%%%%%%%%%%%%%%
%\balancecolumns
\end{document}

%% file: commands.tex
\usepackage{aaai}
\usepackage{times}
\usepackage{helvet}
\usepackage{courier}
\usepackage{savesym}
\usepackage{multirow}
\usepackage{xspace}
\usepackage{graphicx}
\usepackage{url}
\usepackage{enumitem}
\usepackage{color}
\usepackage{amsmath,amssymb,xfrac,amsthm,amsfonts}
\usepackage[tight]{subfigure}
\usepackage{verbatim, xfrac}
\usepackage[titletoc]{appendix}
\setitemize{noitemsep,topsep=1pt,parsep=1pt,partopsep=1pt, leftmargin=12pt}
\setlist{nolistsep}

\newtheorem{theorem}{Theorem}

\makeatletter

\newcommand{\Rmnum}[1]{\expandafter\@slowromancap\romannumeral #1@}
\makeatother
\usepackage{caption}

\usepackage[lined, boxed, ruled, commentsnumbered, noend]{algorithm2e}

\newcommand{\citet}[1]{\citeauthor{#1} (\citeyear{#1})}
\usepackage{algpseudocode}
\DeclareMathOperator*{\argmin}{arg\,min}
\DeclareMathOperator*{\argmax}{arg\,max}

\newcommand{\uexp}{\textsc{UExpSelect}\xspace}
\newcommand{\aexp}{\textsc{AExpSelect}\xspace}
\newcommand{\lucb}{\textsc{Lucb}-1\xspace}
\newcommand{\uniform}{\textsc{Naive}\xspace}

\newcommand{\pushline}{\Indp}% Indent

%% file: introduction.tex
% !TEX root =  hcomp-bestarms-sideinfo.tex
%%%%%%%%%%%%%%%%%%%%%%%%%%%%%%%%%%%%%%%%%%%%%%%%%%%%%%%%%
%%%%%%%%%%%%%%%%%%%%%%%%%%%%%%%%%%%%%%%%%%%%%%%%%%%%%%%%%
\vspace{-3.5mm}
\section{Introduction}\label{sec.introduction}
\vspace{-0.5mm}
The success of a project or a collaborative venture depends critically on acquiring a team of contributors. Beyond increased performance and productivity, hiring a strong team leads to enhanced engagement and retention of workers.

\vspace{0.5mm}
{\it ``A small team of A+ players can run circles around a giant team of B and C players." -- Steve Jobs}
\vspace{0.5mm}

{\bf Crowdsourcing and outsourcing} via online marketplaces further underscores the promise of developing procedures for identifying potential contributors and composing teams. Crowdsourcing and human computation platforms highlight the opportunities for optimizing team building even when a job requester and workers may be half a world apart and have no advance contact. To date, online crowdsourcing markets have largely focused on micro-tasking through enlisting a non-expert crowds of workers, who work independently and contribute to the solution of simple tasks such as performing image annotation and rating web pages. With the increasing complexity of tasks that are crowdsourced, as well as enterprises outsourcing their work, the need to hire skilled workers with an eye to considerations of complementarity and coordinative efforts in a collaboration around problem solving is becoming important.  Contract-based crowdsourcing is another emerging paradigm where workers are recruited on a contract for performing tasks on an ongoing basis.  The online platforms are offering new capabilities to deal with this shift towards expertise-driven crowdsourcing. For instance, oDesk provides opportunities for workers to do self-assessments via the taking of voluntary tests ranging from those evaluating language skills to competencies in more complex disciplines such as programming. The platform provides support for recruiters to conduct interviews and perform online tests for job applicants. Furthermore, most of these marketplaces employ a feedback mechanism that allows task and platform owners to track the skill-specific expertise and reputation of workers to help with future recruiting.

%%%%%%%%%%%%%%%%%%%%%%%%%%%%%%%%%%%%%%%%%%%%%%%%%%%%%%%%%%
\begin{figure*}[t]
\centering
\includegraphics[width=0.7\textwidth]{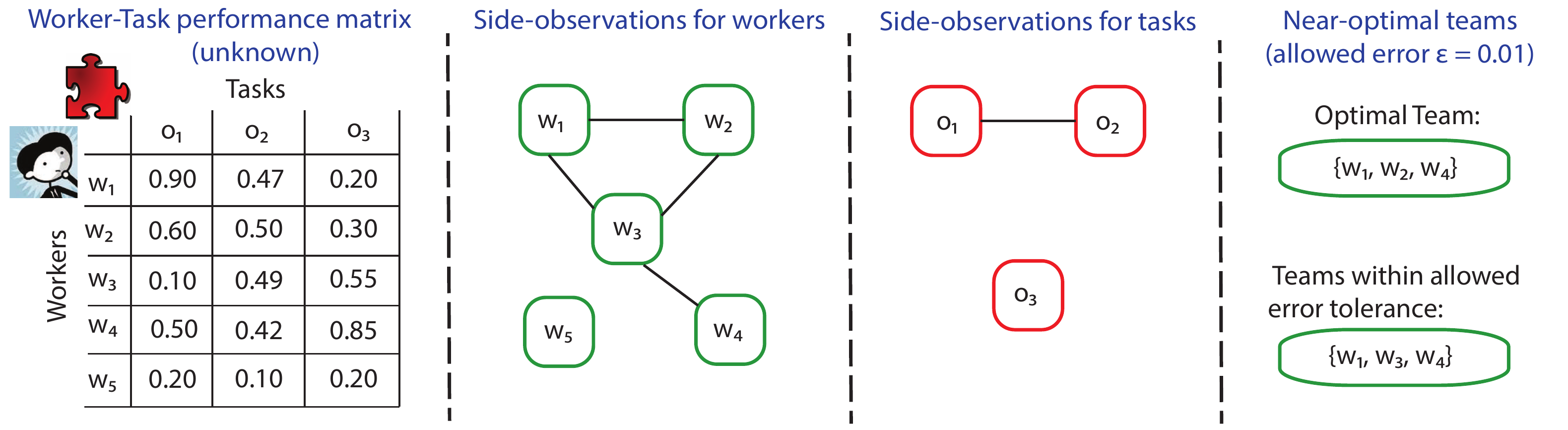}
\vspace{-1mm}
\caption{Illustration of approach on toy example with five workers and three (types of) tasks.}
\label{fig.approach}
\end{figure*}
%%%%%%%%%%%%%%%%%%%%%%%%%%%%%%%%%%%%%%%%%%%%%%%%%%%%%%%%%

{\bf Tasks and the team.} We consider the crowdsourcing setting where the job requester has a predefined heterogeneous set of types of tasks that need to be solved on an ongoing basis. The notion of task types here could alternatively be taken to refer to the unique set of skills that are needed for addressing the needs of a project. For instance, consider an enterprise whose goal is to outsource a project that has three components or categories of tasks, each requiring a particular skill: ({\em i}) web development, ({\em ii}) English to Spanish translation, and  ({\em iii}) video editing. The project would have ongoing assignments of  tasks that would belong to one of these three components. When a new task needs to be executed, it is assigned to the hired team and can be performed by the worker possessing the highest expertise for the skill required for this task. The quality of the hired team could then be quantified by the highest expertise among the team members for each of the skills that are required for this project.

{\bf Learning workers' expertise.}  In the general case, workers' expertise over different types of tasks or the skills is unknown to the recruiter. To learn the worker's expertise for a given type of task, the recruiter can perform an online test or evaluate the performance of the worker via assignment of gold-standard questions for which the ground truth is available. Under standard statistical assumptions, performing more of these tests on a worker would give a better estimate about the expertise level of the worker for a given type of task. The recruiter's goal is to hire a near-optimal team with high likelihood. The main research question is then how to optimally spend the budget (or minimize the total number of tests performed) in order to obtain a sufficiently good estimate of the workers' expertise over all of the required task types and to be able to make the hiring decision under an allowed level of error tolerance.

{\bf Exploiting commonalities.} Typically, the unique number of task types and the total number of job applicants (or the workers that bid for the posted tasks) could be large and hence may require performing large numbers of tests in order to learn the workers' expertise. However, in order to speed up learning, one may be able to exploit the similarities among the tasks and underlying social ties or commonalities among the workers. For instance, consider two types of tasks, requiring skill ``java script" and skill ``ajax". By using group testing, the recruiter may design one test for skill ``java script" that could allow to additionally infer the expertise on skill ``ajax" at no additional cost. Prior knowledge about correlations among workers' expertise and workers'  features (such as demographics) could also be exploited.  Depending on the specific application setting, one may be able to exploit the social ties among workers (or ``participants"). The goal is to design algorithms that can exploit these different kinds of commonalities should they be present.
%%%\ak{?} 
%%%%For example, by designing appropriate tests, one may be able to post the same test to friends' newsfeeds and get feedbacks without any additional cost. 

{\bf Our contributions} can be summarized as follows:
\begin{itemize}
\item We present an algorithmic approach to hiring a team of workers as faced by a recruiter for contract-based crowdsourcing;
\item we provide algorithms with PAC bounds on the required budget to hire a near-optimal team with high confidence. Our algorithms phrase the decision-theoretic problem of team hiring as an instance of online learning for best action selection.
\item We propose a simple model to jointly consider the commonalities among tasks and workers, extend our algorithms to exploit them and
\item evaluate the proposed methods using synthetic data as well as  data collected from the oDesk platform.
\end{itemize}

%%%%%%%%%%%%%%%%%%%%%%%%%%%%%%%%%%%%%%%%%%%%%%%%%%%%%%%%%
%%%%%%%%%%%%%%%%%%%%%%%%%%%%%%%%%%%%%%%%%%%%%%%%%%%%%%%%%

%% file: related.tex
% !TEX root =  hcomp-bestarms-sideinfo.tex
%%%%%%%%%%%%%%%%%%%%%%%%%%%%%%%%%%%%%%%%%%%%%%%%%%%%%%%%%%%%%%%%%%%%%%%%%%
%%%%%%%%%%%%%%%%%%%%%%%%%%%%%%%%%%%%%%%%%%%%%%%%%%%%%%%%%%%%%%%%%%%%%%%%%%
%\vspace{-4mm}
\section{Related Work}\label{sec.related}
%We now give an overview of the related research work.

%%%%%%%%%%%%%%%%%%%%%%%%%%%%%%%%%%%%%%%%%
{\bf Heterogeneous crowdsourcing markets.} 
Our work tackles challenges that arise in heterogeneous crowdsourcing markets where a worker's performance for a given task depends on the required skills and the expertise level of the worker for those skills.  \citet{2013-www_pick-a-crowd} focus on building automated tools to pick the right set of eligible workers for a given task based on the social networking profile of the workers.  \citet{2014-hcomp_matching-crowd} design a mechanism for assigning tasks to workers, under the constraints given in terms of a bipartite graph capturing skills and expertise compatibility of the tasks and workers. Another line of research in these markets involve the study of coordination among workers and formation of teams to perform a desired task. \citet{2010-aaai_generalized-task-markets} introduce the notion of generalized task markets, and how machines and humans can interact together to solve such generalized tasks by forming teams. \citet{2011-hcomp_general_comp_ovw} discuss human computation tasks that require effective coordination among workers, such as itinerary planning or data sorting. Our work presents an algorithmic approach to the challenge of team hiring, with guarantees on optimality of the team and the budget required.

%%%%%%%%%%%%%%%%%%%%%%%%%%%%%%%%%%%%%%%%%
{\bf Learning in crowdsourcing.}
Many problems about learning the performance and characteristics of the crowd can be cast as an instance of online learning with associated explore-exploit dilemma, and hence several solutions use the framework of multi-armed bandits (MAB) \cite{1985_lai_ucb_asymptotically}. \citet{2012_aaai-online_task_assignment} and~\citet{ho2013adaptive} tackle the algorithmic questions concerning learning worker's expertise, task assignment and label inference for heterogeneous classification tasks. However, their goal is to improve the overall prediction accuracy at lower cost, rather different from our work.  \citet{2013-www_truthful-incentives} and \citet{2014-ec_adaptive-contract} consider budgeted variants of MAB for learning the price curve and dynamically adjusting payments based on the quality. 

%%%%%%%%%%%%%%%%%%%%%%%%%%%%%%%%%%%%%%%%%
{\bf Best action selection.}
From a technical perspective, the most similar work to ours is the best action selection problem, a more recently introduced variant of MAB problems \cite{2006-jlmr_explore-1,2009-alt_pure-exploration,2012-icml_pac-subset-selection,chen2014combinatorial}. In these settings, the principal agent explores the problem space (the set of actions or ``arms") for a certain time or budget, commits to a policy of the actions and then exploits.  \citet{2006-jlmr_explore-1} study this model under the PAC (probably approximately correct) setting  \cite{1984-_valiant_pac-theory} and introduce various $(\epsilon, \delta)$-PAC algorithms for best ``arm" identification, {\em i.e.}, provide bounds on the number of samples required to output an $\epsilon$-optimal action with probability at least $(1-\delta)$ using concentration bounds  \cite{hoeffding1963probability}.  \citet{2012-icml_pac-subset-selection} design  an adaptive $(\epsilon, \delta)$-PAC algorithm \lucb for selecting $m$ best actions, using upper and lower confidence bounds. \citet{2014-icml_pac-mab} also study the problem of selecting $m$ best actions, introducing a new aggregate metric and then applying it to the crowdsourcing setting by simulation experiments. The uniform exploration policy introduced by  \citet{2006-jlmr_explore-1} and the adaptive policy  \lucb of  \citet{2012-icml_pac-subset-selection} are the main building blocks of our proposed algorithms. \citet{2011-nips_multi-bandits,2013-icml_multi-best-arm} consider the problem of best arm identification in multiple MAB problem instances by jointly learning over all the problem instances. Our algorithms are also inspired from this idea of jointly identifying best actions over multiple problem instances and we extend the \lucb algorithm to this setting.

%%%%%%%%%%%%%%%%%%%%%%%%%%%%%%%%%%%%%%%%%
{\bf Exploiting commonalities and modeling side-observations.}
A recent line of research has introduced the notion of side-observations to exploit the additional information that can speed up learning. \citet{2011-nips_bandits-to-experts} consider a class of problems that interpolate between bandit feedback and full information settings. They consider the bandit feedback model with side-observations (for instance, such side-observations could arise from user/advertisement similarity, sensor proximity \emph{etc.}) and design algorithms for adversarial settings. \citet{2012-uai_side-info-bandits} and \citet{2014-sigmetrics_mab-social-stochatic} extend the results of the side-observation model for  stochastic settings. Side-observations through correlations \cite{2014-kdd_networked} captures bandit problems where the actions are correlated, as well as pulling one actions invokes these correlated actions accounting for additional rewards and observations, motivated by applications in social advertisement. \citet{2013-nips_gang-bandits} present an algorithm for contextual bandits correlated through an underlying graph. We borrow some of the ideas from \citet{2011-nips_bandits-to-experts} and \citet{2014-sigmetrics_mab-social-stochatic} to exploit the commonalities among tasks and among workers. We present a simple model to jointly consider the commonalities among tasks and workers by representing it as a cross product of two side-observation graphs. Furthermore, for the first time, we apply these side-observation models to the best action selection problem. 

%% file: model.tex
\section{Problem Statement}\label{sec.model}
We now formalize the problem addressed in this paper.

%%%%%%%%%%%%%%%%%%%%%%%%%%%%%%%%%%%%%%%%%%%%%%%%%%%%%%%%%
{\bf Tasks and workers.} We have a set of $M$ types of tasks (simply referred to as tasks henceforth) and $N$ workers (or job applicants) denoted by the sets $\mathcal{O}=\{o_1, o_2,\dots,o_M\}$ and $\mathcal{W}=\{w_1, w_2,\dots,w_N\}$, respectively. We shall assume $N \geq M$, simply meaning that there is at least one unique  job applicant per type of task.  For instance, in Figure~\ref{fig.approach}, we have $M=3$ tasks, and $N=5$ workers.  We model the performance of a worker for a given task as a bounded random variable with unknown mean. Assigning task $o_j \in \mathcal{O}$ to worker  $w_i \in \mathcal{W}$ at time $t$ yields a performance value (as feedback) denoted by random variable $X_{(i,j)}^t$, sampled from an unknown distribution with mean value  $\mu_{(i,j)}$. For simplicity and w.l.o.g, we shall assume that the underlying distribution from which $X_{(i,j)}^t$ is sampled has a bounded support within $[0,1]$. The mean performance values are denoted by an unknown performance matrix $\mu:N \times M \rightarrow \mathcal{R}_{\geq 0}$ with tasks as columns, and workers as rows. We assume a stochastic setting where  $X_{(i,j)}^t$ are \emph{i.i.d.} for any fixed pair of worker $w_i$ and task $o_j$. Also, ${X_{(i,j)}^t}$ are \emph{independent} across $i$, $j$ and $t$.

%%%%%%%%%%%%%%%%%%%%%%%%%%%%%%%%%%%%%%%%%%%%%%%%%%%%%%%%%
{\bf Side-observation model.} The workers and tasks are embedded in some (known) underlying graphs, denoted by $G_w(V_w, E_w)$ and $G_o(V_o,E_o)$. The nodes $V_w \in G_w$ correspond to the $N$ workers, and nodes $V_o \in G_o$ correspond to the $M$ tasks. We shall assume undirected graphs, though the models and results could be extended to the setting of directed graphs as well. The edges in these graphs capture the model of side-observations that may be possible to obtain at no additional cost \cite{2011-nips_bandits-to-experts,2014-sigmetrics_mab-social-stochatic}. In our model, when worker $w_i$ is assigned task $o_j$ at time $t$, apart from observing the performance $X_{(i,j)}^t$, the following additional set of observations become available: 
\vspace{-1mm}
\begin{itemize}
\item $X_{(i,q)}^t \ \forall \ q: \{o_j,o_q\} \in E_o$, the additional observations associated with the tasks neighboring to $o_j$  in $G_o$.
\item $X_{(p,j)}^t \ \forall \ p: \{w_i,w_p\} \in E_w$, the additional observations associated with the workers neighboring to $w_i$  in $G_w$.
\end{itemize}
%\vspace{-1mm}
In Figure~\ref{fig.approach}, assigning task $o_1$ to worker $w_2$ at time $t$ would yield set of observations given by $X^t = \{X_{(2,1)}^t, X_{(2,2)}^t, X_{(1,1)}^t, X_{(3,1)}^t\}$.
%%%would lead to following additional side-observations in our model: we get a sample of the performance of $w_2$ for the task $o_2$, as well as sample of the performance of $w_1$ and $w_3$ for the task $o_1$.  
The goal is to design algorithms that can exploit these side-observations \emph{whenever} present, and smoothly interpolate between the {\em bandit setting} (absence of side-observations, $E_w = E_o = \emptyset$) to the {\em full information setting} (fully connected graphs).
%%%\ak{not quite right -- since you do not observe the cross-product, right?}.  
%%%%\adish{had the intention to say fully connected graph of N*M vertices joined over the cross-product. Perhaps need further clarity.}

%%%%%%%%%%%%%%%%%%%%%%%%%%%%%%%%%%%%%%%%%%%%%%%%%%%%%%%%%
{\bf The objective.}
Our goal is to select or hire a team of workers denoted by $\mathcal{S}^*$, of size at most $M$ from the set $\mathcal{W}$, comprising the highest performing worker for each task $o \in \mathcal{O}$. If the performance matrix is known, the problem is trivial, for instance, in Figure~\ref{fig.approach}, the optimal team is $\{w_1, w_2, w_4\}$. Hence, the goal is to design algorithm that can efficiently learn the performance matrix $\mu[N,M]$ and output a near-optimal team.  In our model, a team $\mathcal{S}$ is  $\epsilon$-optimal, when, for each task $o_j \in \mathcal{O}$, we have:
\vspace{-2mm}
\begin{align}
\forall o_j \in \mathcal{O}, \operatorname*{max}_{w_i \in \mathcal{W}} \mu_{(i,j)} - \operatorname*{max}_{w_i \in \mathcal{S}} \mu_{(i,j)} \leq \epsilon
\end{align}
\vspace{-1.5mm}

In Figure~\ref{fig.approach}, $\{w_1, w_3, w_4\}$ is an $\epsilon$-optimal team for $\epsilon = 0.01$. Given our stochastic assumptions, the algorithm can repeatedly assign a task $o_j$ to worker $w_i$ in order to get a good estimate of the performance $u_{(i,j)}$. We call each such assignment being a {\em test} performed. We assume that each such test poses a unit cost to the algorithm.  We seek algorithms with PAC bounds, \emph{i.e.}, for given positive constants $(\epsilon, \delta)$, the algorithm should output an $\epsilon$-optimal team with probability of at least $(1 - \delta)$.
We measure the efficiency of such a algorithm in terms of the total number of tests required or equivalently the budget spent.
% to output $(\epsilon$-optimal team with probability at least $(1 - \delta)$.

%%%%%%%%%%%%%%%%%%%%%%%%%%%%%%%%%%%%%%%%%%%%%%%%%%%%%%%%%
%%%%%%%%%%%%%%%%%%%%%%%%%%%%%%%%%%%%%%%%%%%%%%%%%%%%%%%%% EXTRA

%% file: algorithm.tex
% !TEX root =  hcomp-bestarms-sideinfo.tex
%%%%%%%%%%%%%%%%%%%%%%%%%%%%%%%%%%%%%%%%%%%%%%%%%%%%%%%%%
%%%%%%%%%%%%%%%%%%%%%%%%%%%%%%%%%%%%%%%%%%%%%%%%%%%%%%%%%
%\vspace{-2mm}
\section{Algorithms for Budgeted Hiring}\label{sec.mechanism}
%We begin by describing the high level ideas behind our algorithms.

%%%%%%%%%%%%%%%%%%%%%%%%%%%%%%%%%%%%%%%%%%%%%%%%%%%%%%%%%
%%%%%%%%%%%%%%%%%%%%%%%%%%%%%%%%%%%%%%%%%%%%%%%%%%%%%%%%%
%\vspace{-1.5mm}
\subsection{Overview of basic approach} \label{sec.mechanism.method1}
To present some of the key insights in designing our algorithms, we first consider a simple setting.
%{\bf Single task ($M=1$) without side observations.}

%%%%%%%%%%%%%%%%%%%%%%%%%%%%%%%%%%%%%%%%%%%%%%%%%%%%%%%%%
%%%%%%%%%%%%%%%%%%%%%%%%%%%%%%%%%%%%%%%%%%%%%%%%%%%%%%%%%
{\bf Single task ($M=1$) without side-observations.} Let us first consider the simple setting of hiring to solve one task, \emph{i.e.}, $M=1$ and the goal is to find an $\epsilon$-optimal worker from set $\mathcal{W}$ with success probability of at least $(1 - \delta)$. We consider the recruiting of team members from among $N$ workers as the set of actions at hand, and reduce the decision problem to the problem of best action selection \cite{2006-jlmr_explore-1,2009-alt_pure-exploration,2012-icml_pac-subset-selection,chen2014combinatorial}.  For example, the $\uniform(\epsilon,\delta)$ algorithm  of \citet{2006-jlmr_explore-1} provides $(\epsilon,\delta)$-PAC guarantees by uniformly allocating a sufficient number of observations for each action to be able to select  $\epsilon$-optimal action with probability at least $(1 - \delta)$. By using Hoeffding's inequality \cite{hoeffding1963probability}, a sufficient number of observations is $\big\lceil \frac{2}{\epsilon^2} \ln(\frac{N}{\delta}) \big\rceil$. This $\uniform(\epsilon,\delta)$ algorithm is the main building block for our proposed algorithm \uexp based on uniform exploration of the actions.

This algorithm is based on uniform exploration and ignores the fact that some actions may be easier or harder to distinguish. For example, in Figure~\ref{fig.approach}, considering task $o_1$, distinguishing $w_5$ from $w_1$ is easier than distinguishing $w_2$ from $w_1$. To tackle this problem,  \citet{2012-icml_pac-subset-selection} design an adaptive $(\epsilon, \delta)$-PAC algorithm \lucb using upper and lower confidence bounds.  \lucb adapts to the complexity of the problem instance, provides distribution dependent bounds and is the state-of-the-art algorithm for the best action selection problem.  We use  \lucb as the main building block for our proposed algorithm \aexp, an adaptive variant of \uexp.

%%%%%%%%%%%%%%%%%%%%%%%%%%%%%%%%%%%%%%%%%%%%%%%%%%%%%%%%%
 %%%%%%%%%%%%%%%%%%%%%%%%%%%%%%%%%%%%%%%%%%%%%%%%%%%%%%%%%
{\bf Multiple tasks ($M>1$) without side-observations.} 
One possible way to tackle this challenge is to consider each task as a separate instance of the best action selection problem, and to use one of the previously discussed algorithms \uniform or \lucb separately. However, one can hope to do better by jointly considering all of the tasks, and allocating the budget across tasks in an adaptive manner.  For instance, in Figure~\ref{fig.approach}, the task $t_2$ is harder than task $t_1$ and task $t_3$ in terms of distinguishing and selecting the best worker. Recently, \citet{2011-nips_multi-bandits} and \citet{2013-icml_multi-best-arm} have addressed this problem of best arm identification in multiple multi-armed bandit (MAB) instances by jointly learning over all of the instances. Our proposed algorithms are inspired from the idea of jointly identifying best workers (the team) for all of the tasks and \aexp extend the \lucb algorithm to this setting.

%%%%%%%%%%%%%%%%%%%%%%%%%%%%%%%%%%%%%%%%%%%%%%%%%%%%%%%%%
%%%%%%%%%%%%%%%%%%%%%%%%%%%%%%%%%%%%%%%%%%%%%%%%%%%%%%%%%
%\vspace{-1.5mm}
\subsection{Exploiting side-observation graphs} \label{sec.mechanism.method2}
Side-observation models \cite{2011-nips_bandits-to-experts,2012-uai_side-info-bandits,2014-sigmetrics_mab-social-stochatic} have been studied mainly in context of regret minimization problems using MAB framework, modeling the observations via an underlying graph connecting the ``arms" of the MAB. Although different ideas have been explored on how to exploit side-observations via an underlying graph, all these ideas revolve around the {\em minimal dominating set} of the side-observation graph $G(V,E)$ denoted by $\texttt{DOM}(G)$. This concept refers to the smallest subset of vertices that cover the rest --- every vertex of the graph $G$ is either in $\texttt{DOM}(G)$ or is directly connected to one of the vertices in $\texttt{DOM}(G)$. 

We extend these ideas to apply the side-observation models for the best action selection problem. Since our proposed algorithms jointly learn over these $M$ tasks, we would like to jointly exploit the side-observation graphs over the tasks and the workers. We can model the side-observation graphs jointly as the cartesian product of two graphs given by $G_w \Box G_o$, denoted as $G_{wo} = (V_{wo}, E_{wo})$. In a cartesian product of graphs, the vertices are given by the cartesian product of the vertex sets of the individual graphs,  $V_{wo} = V_w \times V_o$, or alternatively, $V_{wo} = \{(w_i, o_j): i \in [1 \dots N] \texttt{ and }  j \in [1 \dots M]\}$, \emph{i.e.}, $G_{wo}$ has $M \cdot N$ vertices. The edges are given by $E_{wo}$ such that $\{(w_i, o_j)\}$ and $\{(w_{i'}, o_{j'}\}$ have an edge if either \emph{i}) $w_i = w_{i'}$ and $(o_j, o_{j'}) \in E_o$, or \emph{ii}) $o_j = o_{j'}$ and $(w_i, w_{i'}) \in E_w$.  Let $\gamma(G_{wo})$ denote the minimum size of a dominating set in the resulting graph. Computing the dominating set itself is NP-Hard by a reduction from the set-cover problem \cite{1998-_khuller_cds-algorithmica}. However, an approximate solution can be found of size  upper bounded by $\big(1 + \ln(1 + \texttt{DEG}(G_{wo}))\big)$, where $\texttt{DEG}$  denotes the maximum degree of any vertex in the graph \cite{1998-_khuller_cds-algorithmica}. Let us denote this approximate dominating set as $\widetilde{\texttt{DOM}}(G_{wo})$ and the corresponding approximate dominating number as $\widetilde{ \gamma}_{G_{wo}}$.  

We denote the set of \emph{actions} as $\mathcal{A} = \{a_{(i,j)}: i \in [1 \dots N] \texttt{ and }  j \in [1 \dots M]\}$. Taking action $a_{(i,j)}$ at time $t$ is equivalent to assigning a worker $w_i$ to a task $o_j$ at time $t$. For any action $a_{(i,j)}$, with a slight abuse of notation, we denote its neighboring action belonging to the dominating set as $\widetilde{\texttt{DOM}}(G_{wo}, a_{(i,j)}, \cdot) \in \widetilde{\texttt{DOM}}(G_{wo})$. We call this the {\em dominating action} for $a_{(i,j)}$. For any action  $a_{(i,j)} \in \widetilde{\texttt{DOM}}(G_{wo})$, we denote the set of actions dominated by $a_{(i,j)}$ as $\widetilde{\texttt{DOM}}(G_{wo}, \cdot, a_{(i,j)}) \subseteq V_{wo}$. The main idea used in our algorithms \uexp and \aexp is to replace the picked action $a_{(p,q)}^t$ by its dominating action $a_{(\bar{p},\bar{q})}^t$.

%%%%%%%%%%%%%%%%%%%%%%%%%%%%%%%%%%%%%%%%%%%%%%%%%%%%%%%%%
 %%%%%%%%%%%%%%%%%%%%%%%%%%%%%%%%%%%%%%%%%%%%%%%%%%%%%%%%%
%\vspace{-2.5mm}
\subsection{Model parameters and execution variables} \label{sec.mechanism.method} 
%\vspace{-0.5mm}
We now introduce several model parameters as well as notation that will be useful to describe the algorithms.

%\vspace{-0.5mm}
{\bf $\epsilon$-optimal team.}
 For any task $o_j$, the highest performance among all the workers is given by  $\mu_{(i^*,j)} = \operatorname*{max}_{w_i \in  \mathcal{W}} \mu_{(i,j)}$ and let $w_{(i^*,j)}$  be  the worker with this highest performance. We denote the best worker for this task with the corresponding action $a_{(i^*,j)} \in \mathcal{A}$.  For a given task $o_j$, we can now denote the relative quality of a given worker $w_i$ w.r.t the performance of the best worker available for this task as $\Delta_{(i,j)} = \mu_{(i^*,j)} - \mu_{(i,j)}$. For the specific case of best worker $w_{(i^*,j)}$, this quantity is defined as: $\Delta_{(i^*,j)} = \mu_{(i^*,j)} - \operatorname*{max}_{w_i \in \mathcal{W} \setminus \{w_{(i^*,j)}\}} \mu_{(i,j)}$ denoting the gap with the second best worker for this task. For any task $o_j$, we say a worker $w_i \in \mathcal{W} \setminus \{w_{(i^*,j)}\}$ is $\epsilon$-optimal for $o_j$ if $\Delta_{(i,j)} \leq \epsilon$. We denote this set of  $\epsilon$-optimal workers along with best worker $w_{(i^*,j)}$ as $S_{(\epsilon,j)}$. Now, a solution output $S$ from the algorithm is $\epsilon$-optimal (denoted as $S_{\epsilon}$) if it contains at least one  $\epsilon$-optimal worker for each task, \emph{i.e.}, $ \forall j \in [1 \ldots M]: |S \cap  S_{(\epsilon,j)})| \geq 1$. Putting $\epsilon = 0$ in $S_{\epsilon}$ will correspond to the optimal team.
%%%Our goal is to output such a $S$ of size $M$ that is $\epsilon$-optimal, with at least a probability of $(1 - \delta)$, and minimize the number of tests required to do this. 

%%%%%%%%%%%%%%%%%%%%%%%%%%%%%%%%%%%%%%%%%%%%%%%%%%%%%%%%%
%\vspace{-0.5mm}
{\bf Variables over execution.}
The algorithm will run in time steps, denoted by $t$, where each time step corresponds to the assignment of a task to a worker. Hence, the total number of time steps until execution of the algorithm corresponds to the budget spent or sample complexity for the algorithm. At time step $t$, let $n_{(i,j)}^t$ correspond to the number of times task $o_j$ has been assigned to $w_i$ (or simply, action $a_{(i,j)}$ has been performed). Also, let $y_{(i,j)}^t$ correspond to the number of times total observations have been made about performance of $w_i$ for task $o_j$ (note that, in the absence of side-observations, $n_{(i,j)}^t  = y_{(i,j)}^t$). The current estimate of the mean values are denoted by $\mu_{(i,j)}^t$. With these estimates, we also define $\mu_{(i^*,j)}^t = \operatorname*{max}_{w_i \in  \mathcal{W}} \mu_{(i,j)}^t$, and $w_{(i^*,j)}^t = \operatorname*{arg\,max}_{w_i \in  \mathcal{W}} \mu_{(i,j)}^t$. Similarly, we  define the quantities $\Delta_{(i,j)}^t$ based on current estimate of the performance values $\mu_{(i,j)}^t$.
%%% At time $t$, we can now define the best team $S_{OPT}^t$ to be the one selected based on selecting best workers based on estimates of $\mu_{(i,j)}^t$.  

%%%%%%%%%%%%%%%%%%%%%%%%%%%%%%%%%%%%%%%%%%%%%%%%%%%%%%%%%
\begin{algorithm}[t!]
% \Comment{First batch} 
\nl {\bf Input}: {Tasks: $\mathcal{O}$; Workers: $\mathcal{W}$; Side observation graphs: $G_w$, $G_o$; PAC parameters: $(\epsilon, \delta)$;} \\
\nl {\bf Output}: {Team of workers $S \subseteq W: |S| \leq M$, such that $S$ is $\epsilon$-optimal with probability at least $(1 - \delta)$;} \\
\nl {\bf Initialize}:\\
	\begin{itemize}
    		\item \emph{Compute:} $G_{wo} = G_w \Box G_o$; $\widetilde{\texttt{DOM}}(G_{wo})$;
		\item $t=0$; $S=\emptyset$;
		\item $\forall a_{(ij)} \in A$: $\mu_{(i,j)}^t = 0$; $n_{(i,j)}^t = 0$; $y_{(i,j)}^t = 0$;
			  %$\Delta_{(i,j)}(t)=0$; 
    \end{itemize}
%\Begin{
\While { $\exists \ a_{(i,j)} \in \mathcal{A}: y_{(i,j)}^t < \big\lceil \frac{2}{\epsilon^2} \ln(\frac{M \cdot N}{\delta}) \big\rceil$}{
\nl	$a_{(p,q)}^t = \argmin_{a_{(i,j)} \in \mathcal{A}} y_{(i,j)}^t$; \Comment{Greedy action}\label{alg.uexp.1} \\
% \Comment{Action with minimal observations} 
\nl	$a_{(\bar{p},\bar{q})}^t = \widetilde{\texttt{DOM}}(G_{wo}, a_{(p,q)}^t, \cdot)$;\label{alg.uexp.2}  \\
% \Comment{Dominating action for $a_{(p,q)}^t$} 
\nl	{\bf Perform action:} $a_{(\bar{p},\bar{q})}^t$; \Comment{Assign $o_{\bar{q}}$ to $w_{\bar{p}}$} \\
\nl	{\bf Feedback:} Obtain observations $X^t$ for actions dominated by $a_{(\bar{p},\bar{q})}^t$: $\widetilde{\texttt{DOM}}(G_{wo}, \cdot, a_{(\bar{p},\bar{q})}^t)$ \;
\nl {\bf Update Variables:}\\
    \begin{itemize}
    \item $n_{(\bar{p},\bar{q})}^{t+1} = n_{(\bar{p},\bar{q})}^t + 1$;
    \item $\forall a_{(i,j)} \in \widetilde{\texttt{DOM}}(G_{wo}, \cdot, a_{(\bar{p},\bar{q})}^t)$, $y_{(i,j)}^{t+1} = y_{(i,j)}^t + 1$;
    \item $\forall a_{(i,j)} \in \widetilde{\texttt{DOM}}(G_{wo}, \cdot, a_{(\bar{p},\bar{q})}^t), \texttt{update } \mu_{(i,j)}^{t+1}$  from $X^t$;
    \item $t = t+1$;
    \end{itemize}
}
\ForEach {$j \in [1 \ldots M]$}{
\nl	$w_{(i^*,j)}^t = \argmax_{w_i \in  \mathcal{W}}  \mu_{(i,j)}^t$;\\
\nl	$S \leftarrow S \cup \{w_{(i^*,j)}^t\}$; \\
}
\nl {\bf Output}: $S$\\
%}
\caption{Algorithm \uexp}
\label{alg.uexp} 
\end{algorithm}
%
%%%%%%%%%%%%%%%%%%%%%%%%%%%%%%%%%%%%%%%%%%%%%%%%%%%%%%%%%

%%%%%%%%%%%%%%%%%%%%%%%%%%%%%%%%%%%%%%%%%%%%%%%%%%%%%%%%%
%%%%%%%%%%%%%%%%%%%%%%%%%%%%%%%%%%%%%%%%%%%%%%%%%%%%%%%%%
%\vspace{-2.5mm}
\subsection{Algorithm \uexp} \label{sec.mechanism.uexp}
%\vspace{-0.5mm}
%Static Exploration
%\subsection{Static Exploration {UE{\small \textbf{XP}}S{\small \textbf{ELECT}}}} \label{sec.mechanism.uexp}

We now present our first algorithm \uexp, shown in Algorithm~\ref{alg.uexp},  based on the uniform exploration of all the actions extending ideas of \uniform algorithm \cite{2006-jlmr_explore-1}. At each iteration, the algorithm selects the action $a_{(p,q)}^t$ with minimal number of observations $y_{(p,q)}^t$ (Step~\ref{alg.uexp.1}). This choice is natural and can be thought of as ``greedy" in order to quickly move towards termination of the algorithm. Given the side-observation model, algorithm takes the action $a_{(\bar{p},\bar{q})}^t$  (Step~\ref{alg.uexp.2}), \emph{i.e.}, the one that dominates $a_{(p,q)}^t$, as taking  $a_{(\bar{p},\bar{q})}^t$ also gives us the desired observation needed for $a_{(p,q)}^t$. Then, it receives the observation set  $X^t$, corresponding to all the actions that are dominated by $a_{(\bar{p},\bar{q})}^t$, and updates the corresponding variables. Once every action has made observations of at least  $\big\lceil \frac{2}{\epsilon^2} \ln(\frac{M \cdot N}{\delta}) \big\rceil$, the algorithm selects the best set  $S$ based on the observed performances $\mu_{(i,j)}^t$. Note that, if we ignore the side-observation model, then $a_{(\bar{p},\bar{q})}^t \equiv a_{(p,q)}^t$, and the observations set correspond to singleton set, given by $X^t = \{X_{(p,q)}^t\}$.
%%%, \emph{i.e.}, for each task $o_j$, it selects the best worker $w_{(i^*,j)}^t$ and finally outputs the team set $S$

%%%%%%%%%%%%%%%%%%%%%%%%%%%%%%%%%%%%%%%%%%%%%%%%%%%%%%%%%
%%%%%%%%%%%%%%%%%%%%%%%%%%%%%%%%%%%%%%%%%%%%%%%%%%%%%%%%%

%%%%%%%%%%%%%%%%%%%%%%%%%%%%%%%%%%%%%%%%%%%%%%%%%%%%%%%%%
\begin{algorithm}[t!]
% \Comment{First batch} 
\nl {\bf Input}: {Tasks: $\mathcal{O}$; Workers: $\mathcal{W}$; Side observation graphs: $G_w$, $G_o$; PAC parameters: $(\epsilon, \delta)$;} \\
\nl {\bf Output}: {Team of workers $S \subseteq W: |S| \leq M$, such that $S$ is $\epsilon$-optimal with probability at least $(1 - \delta)$;} \\
\nl {\bf Initialize}:\\
	\begin{itemize}
    		\item \emph{Compute:} $G_{wo} = G_w \Box G_o$; $\widetilde{\texttt{DOM}}(G_{wo})$;
		\item $t=0$; $S^t=\emptyset$;  $R^t=\mathcal{O}$;
		\item $\forall a_{(ij)} \in \mathcal{A}$: $\mu_{(i,j)}^t = 0$; $n_{(i,j)}^t = 0$; $y_{(i,j)}^t = 0$;
		\item $\forall a_{(ij)} \in \mathcal{A}$: $\beta_{(i,j)}^t \rightarrow \infty$; $\forall o_{j} \in \mathcal{O}$: $\Delta_j^t \rightarrow \infty$;		
    \end{itemize}
%\Begin{
\While {$R^t \neq \emptyset$}{
\nl	$o_q^t = \argmax_{o_j \in R^t} \Delta_j^t$;\label{alg.aexp.1} \\
% \Comment{Task with maximum uncertainity}
\nl $w_{(i^*,q)}^t = \argmax_{w_i \in \mathcal{W}} \mu_{(i,q)}^t$; \label{alg.aexp.2}\\
\nl $w_{(i^\bullet,q)}^t = \argmax_{w_i \in \mathcal{W} \setminus \{w_{(i^*,q)}^t\}} \big(\mu_{(i,q)}^t + \beta(y_{(i,q)}^t, t)\big)$; \label{alg.aexp.3}\\
\nl $w_p^t = \argmax_{w_i \in \{w_{(i^*,q)}^t, w_{(i^\bullet,q)}^t\}}   \beta(y_{(i,q)}^t, t)$; \label{alg.aexp.4} \\
% \Comment{Worker with maximum uncertainity}
\nl	$a_{(p,q)}^t \leftarrow (w_p^t, o_q^t)$; \Comment{Greedy action} \label{alg.aexp.5} \\
\nl	$a_{(\bar{p},\bar{q})}^t = \widetilde{\texttt{DOM}}(G_{wo}, a_{(p,q)}^t, \cdot)$; \\
%\Comment{Dominating action for $a_{(p,q)}^t$}
\nl	{\bf Perform action:} $a_{(\bar{p},\bar{q})}^t$; \Comment{Assign $o_{\bar{q}}$ to $w_{\bar{p}}$} \\
\nl	{\bf Feedback:} Obtain observations $X^t$ for actions dominated by $a_{(\bar{p},\bar{q})}^t$: $\widetilde{\texttt{DOM}}(G_{wo}, \cdot, a_{(\bar{p},\bar{q})}^t)$ \;
\nl {\bf Update Variables:}\\
    \begin{itemize}
    \item  $n_{(\bar{p},\bar{q})}^{t+1} = n_{(\bar{p},\bar{q})}^t + 1$;
    \item $\forall a_{(i,j)} \in \widetilde{\texttt{DOM}}(G_{wo}, \cdot, a_{(\bar{p},\bar{q})}^t)$, $y_{(i,j)}^{t+1} = y_{(i,j)}^t + 1$;
	\item $\forall a_{(i,j)} \in \widetilde{\texttt{DOM}}(G_{wo}, \cdot, a_{(\bar{p},\bar{q})}^t), \texttt{update }\mu_{(i,j)}^{t+1}$  from $X^t$;    
    \item $t = t+1$;
    \item $\forall o_j \in R^t, \texttt{ update } \Delta_j^t$;
    \end{itemize}
\nl {\bf Update Solution:}\\
	{
		\pushline
		\ForEach {$o_j \in R^t$}{
			\If {$\Delta_j^t \leq \epsilon$}{
\nl				$w_{(i^*,j)}^t = \argmax_{w_i \in \mathcal{W}} \mu_{(i,j)}^t$;\\
\nl              $S^t = S^t \cup \{w_{(i^*,j)}^t\}$; \\
\nl				$R^t = R^t \setminus \{o_j\}$; \\
			}
		}
	}
}
\nl {\bf Output}: $S^t$\\
%}
\caption{Algorithm \aexp}
\label{alg.aexp} 
\end{algorithm}
%
%\ForEach {$j \in [1 \ldots M]$}{
%\nl	$w^*(j,t) = \argmax_{i \in [1 \ldots N]}  \mu_{i,j}(t)$;\\
%\nl	$S \leftarrow S \cup \{w^*(j,t)\}$; \\
%}
%%%%%%%%%%%%%%%%%%%%%%%%%%%%%%%%%%%%%%%%%%%%%%%%%%%%%%%%%

%%%%%%%%%%%%%%%%%%%%%%%%%%%%%%%%%%%%%%%%%%%%%%%%%%%%%%%%%
%%%%%%%%%%%%%%%%%%%%%%%%%%%%%%%%%%%%%%%%%%%%%%%%%%%%%%%%%
%\vspace{-2mm}
\subsection{Algorithm \aexp} \label{sec.mechanism.aexp}
%\vspace{-0.5mm}
In order to adapt the algorithm to the variability of the hardness of the problem in identifying suboptimal workers across tasks and within one given task, we present a second algorithm \aexp, based on ideas of \lucb algorithm \cite{2012-icml_pac-subset-selection}. In order to present \aexp, we introduce some specific terminology as well as the approach used to pick the actions. 

First, we associate confidence bounds, \emph{i.e.}, a high probability bound over the estimates of the performance $\mu_{(i,j)}^t$. This is denoted by the function $\beta(y_{(i,j)}^t, t)$. The specific form of function we use, as used in \lucb, is given by $\beta(y, t) = \sqrt{\frac{1}{2 \cdot y} \ln{\Big(\frac{5}{4} \cdot \frac{M \cdot N}{\delta} \cdot {t^4}\Big)}}$. One of the key intuitions behind this specific function is that we seek to ensure that the probability of the event that the confidence interval bounds are \emph{ever} violated over the \emph{lifespan} of the algorithm is bounded by $\delta$. For a given action $a_{(i,j)}$,  the upper and lower confidences over the performance estimate $\mu_{(i,j)}^t$ are given as $\big(\mu_{(i,j)}^t + \beta(y_{(i,j)}^t, t)\big)$ and  $\big(\mu_{(i,j)}^t - \beta(y_{(i,j)}^t, t)\big)$ respectively.  

At a given time $t$ and for a given task $o_j$, we denote the worker with highest empirically observed performance as $w_{(i^*,j)}^t$, given by: 
\begin{align}
w_{(i^*,j)}^t= \argmax_{w_i \in \mathcal{W}} \mu_{(i,j)}^t
\end{align}
Next, from the remaining $N - 1$ workers, we find the worker  with maximum value of upper confidence of performance estimate as follows: 
\begin{align}
w_{(i^\bullet,j)}^t = \argmax_{w_i \in \mathcal{W} \setminus \{w_{(i^*,j)}^t\}} \big(\mu_{(i,j)}^t + \beta(y_{(i,j)}^t, t)\big)
\end{align}

The empirical mean of $w_{(i^*,j)}^t$ is denoted by $\mu_{(i^*,j)}^t$,  and  has lower confidence bound of $\big(\mu_{(i^*,j)}^t - \beta(y_{(i^*,j)}^t,t)\big)$.  For  $w_{(i^\bullet,j)}^t$, the empirical mean is denoted by $\mu_{(i^{\bullet},j)}^t$,  and  has upper confidence bound of $\big(\mu_{(i^{\bullet},j)}^t + \beta(y_{(i^{\bullet},j)}^t,t)\big)$. The quantity that is of particular interest is the gap between upper confidence bound on $\mu_{(i^{\bullet},j)}^t$ and lower confidence bound on  $\mu_{(i^*,j)}^t$. Intuitively, as we get increasing numbers of observations and confidence widths shrink, this gap should reduce to below zero. We denote this quantify for task $o_j$ as follows:
\begin{align}
\Delta_j^t = \big(\mu_{(i^{\bullet},j)}^t + \beta(y_{(i^{\bullet},j)}^t,t)\big) - \big(\mu_{(i^*,j)}^t - \beta(y_{(i^*,j)}^t,t)\big) \label{eq.stoptask}
\end{align}

Based on the ideas from \lucb, the algorithm can commit to worker $w_{(i^*,j)}^t$ for task $o_j$ whenever $\Delta_j^t \leq \epsilon$, and this is $\epsilon$-optimal choice, as long as the confidence intervals are not violated. Intuitively, we are taking the worst-case estimate of $w_{(i^*,j)}^t$ and highest of the best-case estimate from the remaining workers --- ensuring this difference being less than $\epsilon$ is sufficient to commit to worker $w_{(i^*,j)}^t$. 
%%%This will be sufficient for us to have the desired PAC bounds across all the $M$ tasks. 

\aexp is shown in Algorithm~\ref{alg.aexp}. At each iteration, the algorithm first selects the task with highest $\Delta_j^t$ denoted by index $o_q^t$ (Step~\ref{alg.aexp.1}). Then, it finds the corresponding workers $w_{(i^*,q)}^t$ and $w_{(i^\bullet,q)}^t$  (Step~\ref{alg.aexp.2}, ~\ref{alg.aexp.3}). Then, the greedy choice of action $a_{(p,q)}^t$  is based on choosing the worker with higher confidence width among $w_{(i^*,q)}^t$ and $w_{(i^\bullet,q)}^t$  (Step~\ref{alg.aexp.4}, ~\ref{alg.aexp.5}). Note that, the solution set $S^t$ is built over time. The algorithm maintains a set of tasks $R^t$ as the tasks for which a worker still needs to be selected. As soon as the condition in Equation~\ref{eq.stoptask} is met for a task, that task is no longer considered for further actions and removed from $R^t$. The algorithm terminates when $R^t$ is empty. Note that there is a common time clock across all the tasks. Jointly learning over all the tasks ensures that the algorithm can allocate more assignments for the tasks which have maximum uncertainty. Furthermore, it allows us to jointly exploit the side-observation graphs.

%%%%%%%%%%%%%%%%%%%%%%%%%%%%%%%%%%%%%%%%%%%%%%%%%%%%%%%%%
%%%%%%%%%%%%%%%%%%%%%%%%%%%%%%%%%%%%%%%%%%%%%%%%%%%%%%%%%

%% file: analysis.tex
%%%%%%%%%%%%%%%%%%%%%%%%%%%%%%%%%%%%%%%%%%%%%%%%%%%%%%%%%
%%%%%%%%%%%%%%%%%%%%%%%%%%%%%%%%%%%%%%%%%%%%%%%%%%%%%%%%%
%\vspace{-3mm}
\section{Performance Analysis}\label{sec.analysis}
We now analyze the performance of the proposed algorithms \uexp and \aexp. Most of the results below can be derived using the proof techniques of  \uniform  \cite{2006-jlmr_explore-1} and \lucb 
\cite{2012-icml_pac-subset-selection}, and can be seen as extension of their results.

%%%%%%%%%%%%%%%%%%%%%%%%%%%%%%%%%%%%%%%%%%%%%%%%%%%%%%%%%
%\vspace{-1mm}
\subsection{Performance Bounds for \uexp}

Let us consider the case of the absence of side-observations, which is equivalent to setting $E_w = \emptyset$ and $E_o = \emptyset$. In this case, $a_{(\bar{p},\bar{q})}(t) \equiv a_{(p,q)}^t$, observations set correspond to singleton set of $X^t = \{X_{(p,q)}^t\}$, and $\widetilde{\texttt{DOM}}(G_{wo})$ and $\texttt{DOM}(G_{wo})$ are both equal to $V_{wo}$. In fact, in terms of performance bounds, the algorithm \uexp can be seen as equivalent to running $M$ instances of \uniform-$(\epsilon, \frac{\delta}{M})$.  Based on Theorem~6 from \citet{2006-jlmr_explore-1}, the sample complexity of  \uniform-$(\epsilon,\delta)$ for one instance of the problem with $N$ actions is given by $\Big(N \cdot  \Big\lceil \frac{2}{\epsilon^2} \ln(\frac{N}{\delta}) \Big\rceil \Big)$. Hence, the sample complexity of  \uexp  in the absence of side-observations is given by $\Big(M \cdot N \cdot  \Big\lceil \frac{2}{\epsilon^2} \ln(\frac{M \cdot N}{\delta}) \Big\rceil \Big)$. The PAC-$(\epsilon, \delta)$ guarantees hold simply from the correctness of  \uniform-$(\epsilon,\delta)$. The fact that we ran $M$ instances of  \uniform with $\frac{\delta}{M}$ ensures that the error probability is bounded by $\delta$ using the union bound. Next, we can state the improvement in performance obtained by accounting for side-observations in Theorem~\ref{theorem.uexp}.

%\vspace{-1mm}
\begin{theorem}\label{theorem.uexp}
The algorithm \uexp is $(\epsilon, \delta)$-PAC optimal with sample complexity of $ \Big(\widetilde{ \gamma}_{G_{wo}} \cdot  \Big\lceil \frac{2}{\epsilon^2} \ln(\frac{M \cdot N}{\delta}) \Big\rceil \Big)$, where $G_{wo} = G_w \Box G_o$ and $\widetilde{ \gamma}_{G_{wo}} \leq \big(1 + \ln(1 + \texttt{DEG}(G_{wo}))\big) \cdot \gamma_{G_{wo}}$.
\end{theorem}
%\vspace{-1mm}

Recall that $\widetilde{\texttt{DOM}}(G_{wo})$ denotes the polynomial-time approximation of the dominating set for $G_{wo}$ and has size bounded by  $\big(1 + \ln(1 + \texttt{DEG}(G_{wo}))\big)$  \cite{1998-_khuller_cds-algorithmica}. By taking each action of $\widetilde{\texttt{DOM}}(G_{wo})$ once, the entire set of actions is covered. Hence by taking $\widetilde{ \gamma}_{G_{wo}}$ actions, we get observations of the $M \cdot N$ actions resulting in a potential saving of tests by factor of $\frac{\widetilde{ \gamma}_{G_{wo}}}{M \cdot N}$. Importantly, the greedy way of selecting the actions in  Step~\ref{alg.uexp.1} of Algorithm~\ref{alg.uexp} ensures that all of the actions in $\widetilde{ \texttt{DOM}}(G_{wo})$ are scanned uniformly.

%%%%%%%%%%%%%%%%%%%%%%%%%%%%%%%%%%%%%%%%%%%%%%%%%%%%%%%%%
%\vspace{-2mm}
\subsection{Performance Bounds for \aexp}

Let us define $\Delta_{(i,j, \alpha)} = \max_{} \{\Delta_{(i,j)}, \alpha\}$, for any value of $\alpha_{\geq 0}$. In particular, we are interested in quantities $\Delta_{(i,j, \sfrac{\epsilon}{2})}$. Let us again begin by considering the case of the absence of side-observations. One way to tackle this problem is then to run  $M$ instances of \lucb-$(\epsilon, \frac{\delta}{M})$ algorithm, each with its own time clock. Based on Theorem~6 from \citet{2012-icml_pac-subset-selection}, the expected sample complexity of  \lucb-$(\epsilon,\delta)$ for one instance of the problem with $N$ actions for a particular task $o_j$ is given by: 

%\vspace{-2mm}
\begin{align}
O\Big(\big(\sum_{i \in [N]} \frac{1}{\Delta_{(i,j, \frac{\epsilon}{2})}^2} \big) \ln\big(\frac{1}{\delta} \cdot \sum_{i \in [N]} \frac{1}{\Delta_{(i,j, \frac{\epsilon}{2})}^2} \big) \Big) \label{eq.lucb.1}
%\vspace{-1mm}
\end{align}

The expected sample complexity of running $M$ instances of \lucb-$(\epsilon, \frac{\delta}{M})$ is then given by:

%\vspace{-2mm}
\begin{align}
O\bigg( \sum_{j \in [M]} \Big(\big(\sum_{i \in [N]} \frac{1}{\Delta_{(i,j, \frac{\epsilon}{2})}^2} \big) \ln\big(\frac{M}{\delta} \cdot \sum_{i \in [N]} \frac{1}{\Delta_{(i,j, \frac{\epsilon}{2})}^2} \big) \Big)\bigg) \label{eq.lucb.2}
%\vspace{-1mm}
\end{align}

However, by jointly learning across all the tasks, an algorithm can adaptively allocate assignments across the tasks. \aexp is based on this idea, originally proposed in  \citet{2011-nips_multi-bandits,2013-icml_multi-best-arm} and it extends \lucb algorithm to this joint setting. Intuitively, the main reason this is possible in the best-action selection problems is because the problem complexity is defined in terms of relative ``gap" $\Delta_{(i,j)}$ which can be mixed together for all the tasks $o_j$ to create one pool of $M \cdot N$ actions defined by their correspond gaps $\Delta_{(i,j)}$. Then, by using a common time clock over these  $M \cdot N$ actions, the main technical results of \lucb extends to this joint setting \cite{2013-icml_multi-best-arm}. The sample complexity of \aexp is given in Theorem~\ref{theorem.aexp}, which is based on Theorem~6 from \cite{2012-icml_pac-subset-selection}.

\begin{theorem}\label{theorem.aexp}
The algorithm \aexp is $(\epsilon, \delta)$-PAC optimal with expected sample complexity given by
$$O\bigg(\Big(\sum_{j \in [M]} \sum_{i \in [N]} \frac{1}{\Delta_{(i,j, \frac{\epsilon}{2})}^2} \Big) \ln\Big(\frac{1}{\delta} \cdot \sum_{j \in [M]} \sum_{i \in [N]} \frac{1}{\Delta_{(i,j, \frac{\epsilon}{2})}^2} \Big) \bigg)$$
\end{theorem}

Note that the above sample complexity bound is similar in structure as given in Equation~\ref{eq.lucb.1} with total of $M \cdot N$. However, this is different compared to one obtained in  Equation~\ref{eq.lucb.2} by running  $M$ instances of \lucb-$(\epsilon, \frac{\delta}{M})$ with separate time clock for each task. In fact, when all the tasks are of equal hardness defined by quantity $ \sum_{i \in [N]} \frac{1}{\Delta_{(i,j, \frac{\epsilon}{2})}^2}$ for a given $o_j$, the sample complexity in Theorem~\ref{theorem.aexp} and Equation~\ref{eq.lucb.2} is same. 

This bound in  Theorem~\ref{theorem.aexp} is loose in the sense that it doesn't explicitly account for the performance gain achieved by the side-observations, even though \aexp uses the same approach as that used in \uexp to exploit side-observation graphs.  In the worst-case, the static model of side-observations (\emph{i.e.}, a pre-computed and fixed dominating set) that we used, does not help boost the performance for an adaptive algorithm. Intuitively, and as we observed during empirical evaluations, for the problem instances that are uniformly difficult, we tend to gain more value from side-observations. However in such cases, \aexp tend to behave more closely as \uexp. For more skewed tasks and workers in terms of difficultly and performance, the ``easier" to identify workers and tasks gets ``eliminated" over time, and hence the value of side-observations diminish as well.  Hence, for adaptive algorithms like \aexp, a more effective way of exploiting side-observations would need policies that construct dynamic dominating sets at every time step taking into account the remaining uncertainties over the actions.
%%%%%%%%%%%%%%%%%%%%%%%%%%%%%%%%%%%%%%%%%%%%%%%%%%%%%%%%%
%%%%%%%%%%%%%%%%%%%%%%%%%%%%%%%%%%%%%%%%%%%%%%%%%%%%%%%%% EXTRA
%\subsection{Performance Bounds for {AE{\normalsize \textbf{XP}}S{\normalsize \textbf{ELECT}}}}
%%Recall that we defined. For a given task $o_j$, we can now denote the quality of a given worker $w_i$ w.r.t the performance of the best worker available for this task. 
%%Recall that we defined $\Delta_{(i,j)} = \mu_{(i^*,j)} - \mu_{(i,j)}$. 

%% file: experiments.tex
%%%%%%%%%%%%%%%%%%%%%%%%%%%%%%%%%%%%%%%%%%%%%%%%%%%%%%%%%
%%%%%%%%%%%%%%%%%%%%%%%%%%%%%%%%%%%%%%%%%%%%%%%%%%%%%%%%%
\section{Experimental Evaluation}\label{sec.experiments}
We now report on the results of our experiments.

%%%%%%%%%%%%%%%%%%%%%%%%%%%%%%%%%%%%%%%%%%%%%%%%%%%%%%%%%
\begin{figure*}[t!]
\centering
   \subfigure[Precision with increasing budget]{
     \includegraphics[width=0.32\textwidth]{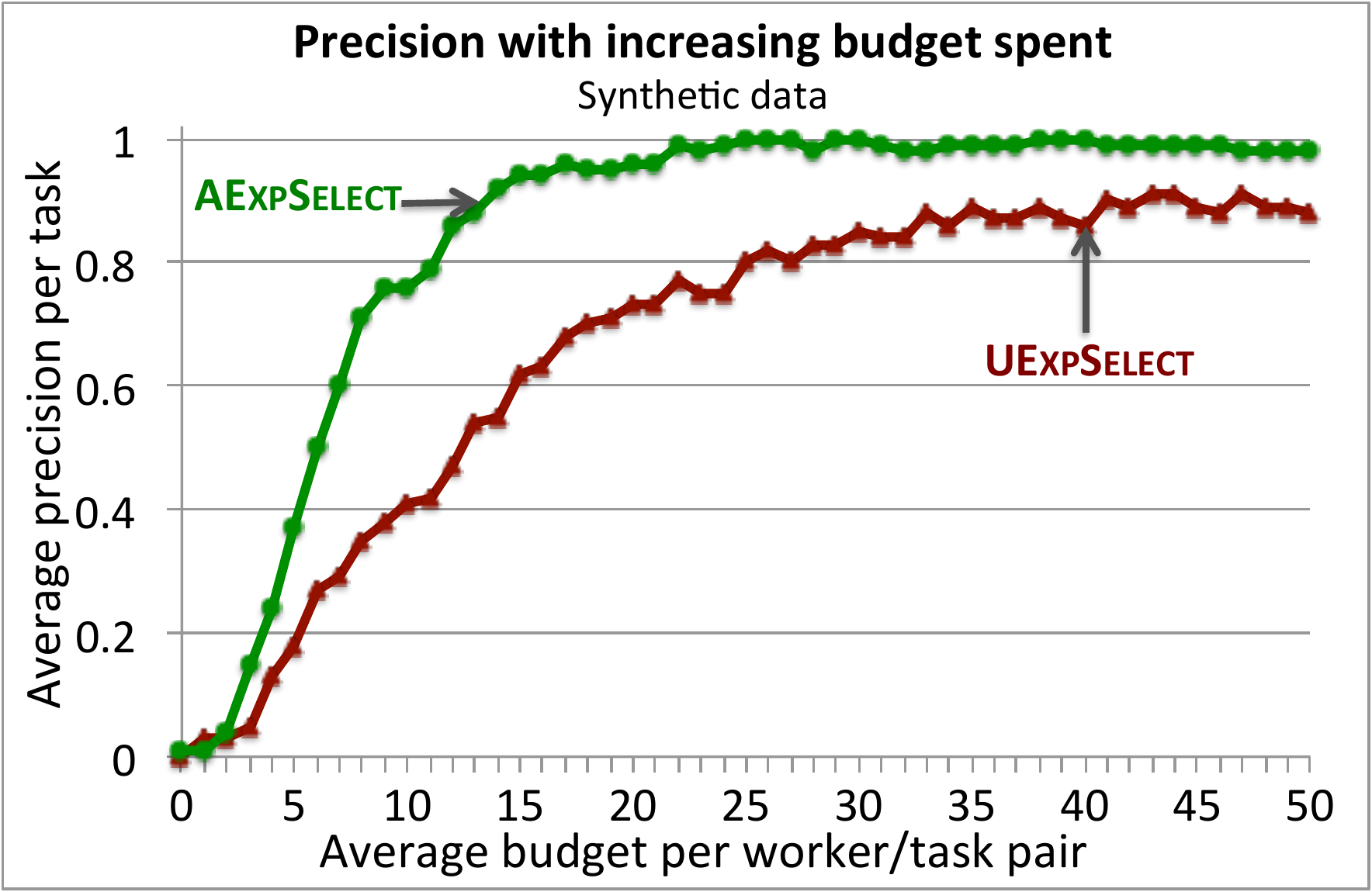}
     \label{fig.syn1}
   }
   \subfigure[Performance gap with increasing budget]{
     \includegraphics[width=0.32\textwidth]{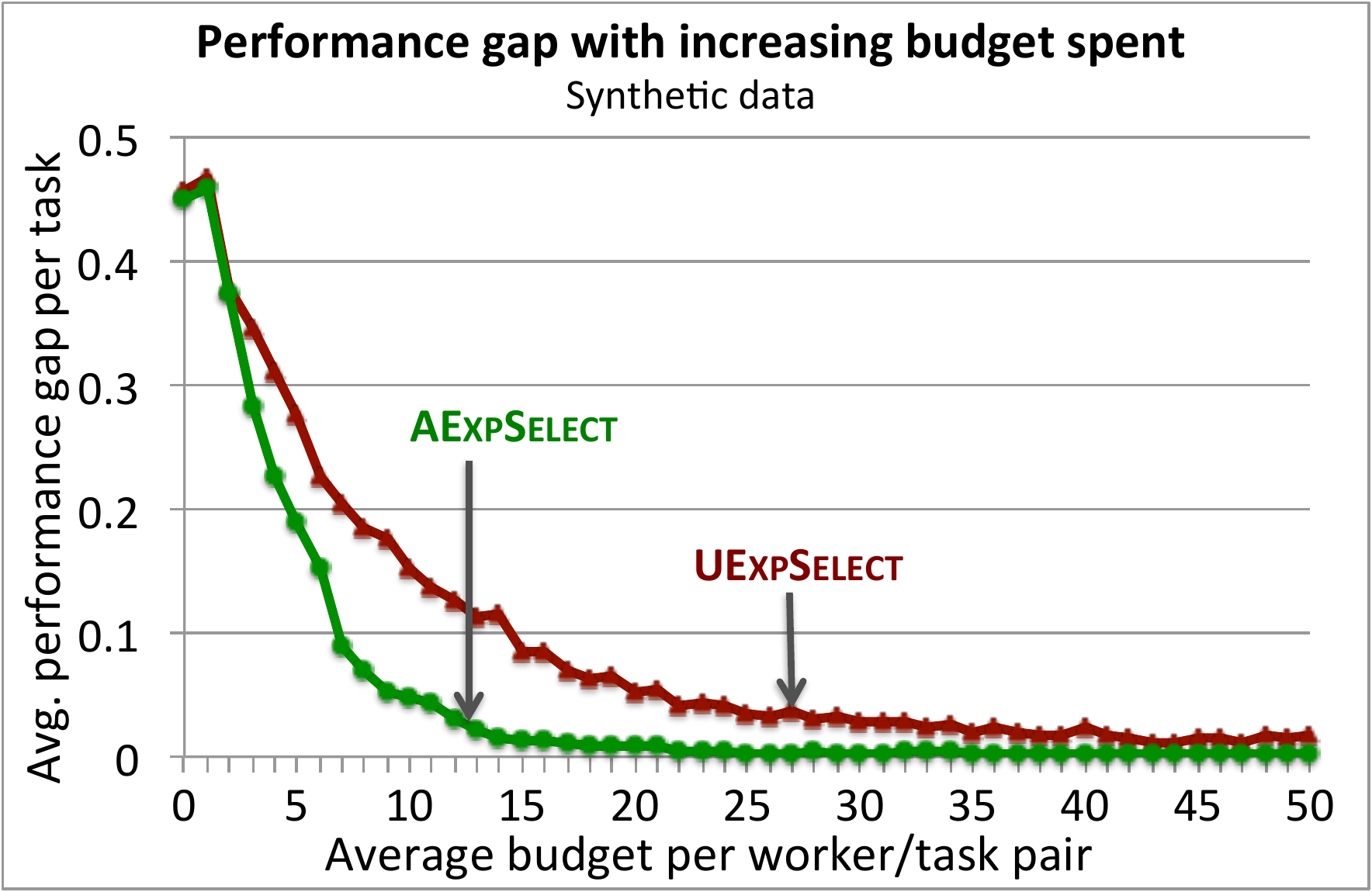}
     \label{fig.syn2}
   }
   \subfigure[Decreasing average gap $\Delta_j^{min}$ in data]{
     \includegraphics[width=0.32\textwidth]{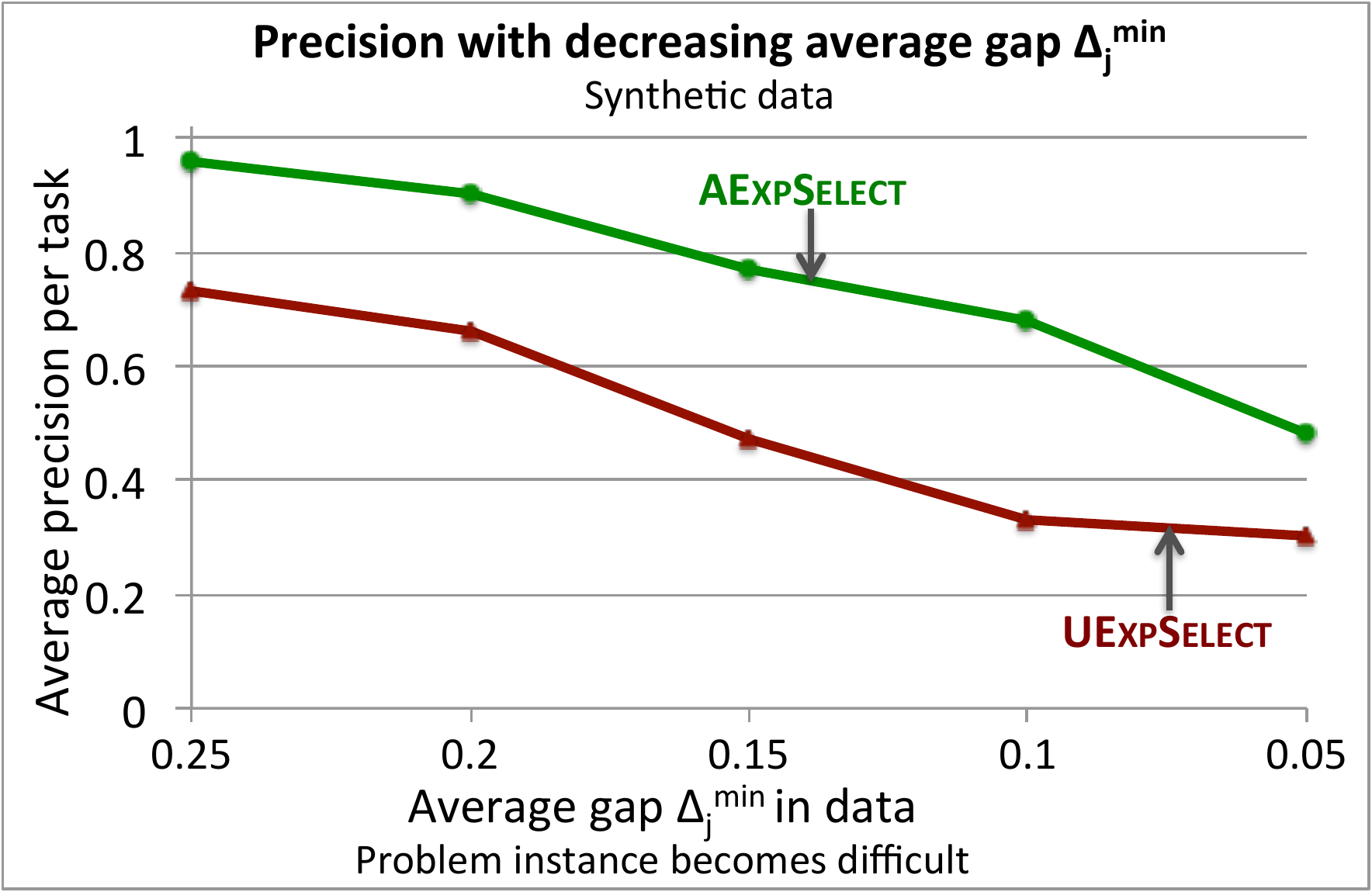}
    \label{fig.syn3}
   }
\vspace{-1mm}
\caption{Experimental results on synthetic data with absence of side-observation graphs. In Figure~\ref{fig.syn1},\ref{fig.syn2}, the budget is varied, and the metrics of average precision and average performance gap are measured, respectively.  In Figure~\ref{fig.syn3}, budget is kept fixed to $20$ per worker/task pair, and average $\Delta_{j}^{min}$ is changed from $0.25$ to $0.05$, making the problem instance difficult.} 
\label{fig.resuts1}
\end{figure*}

%%%%%%%%%%%%%%%%%%%%%%%%%%%%%%%%%%%%%%%%%%%%%%%%%%%%%%%%%
\begin{figure*}[t!]
\centering
   \subfigure[Without side-observations]{
     \includegraphics[width=0.32\textwidth]{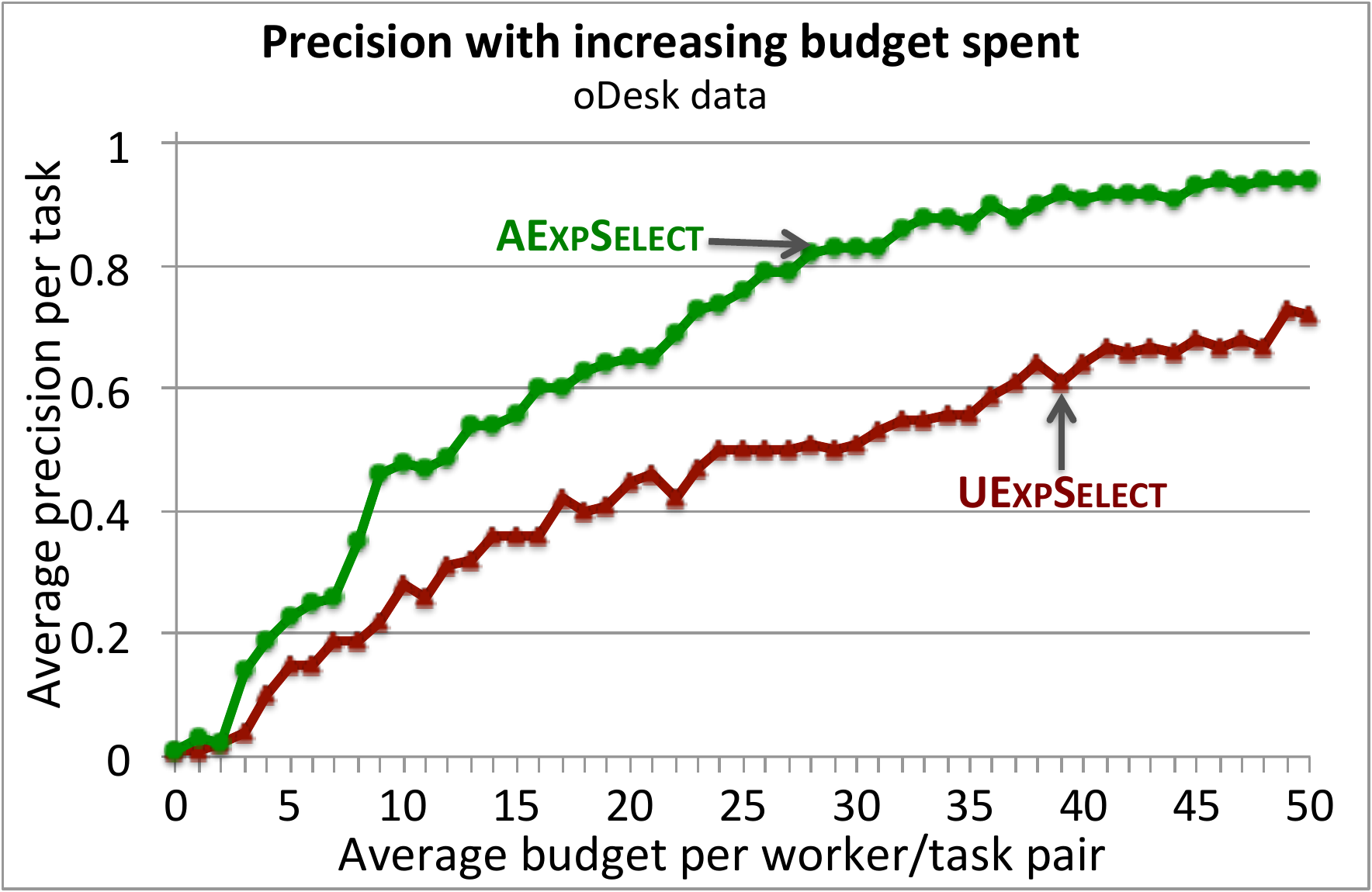}
     \label{fig.odesk1}
   }
   \subfigure[Side-observations over tasks]{
    \includegraphics[width=0.32\textwidth]{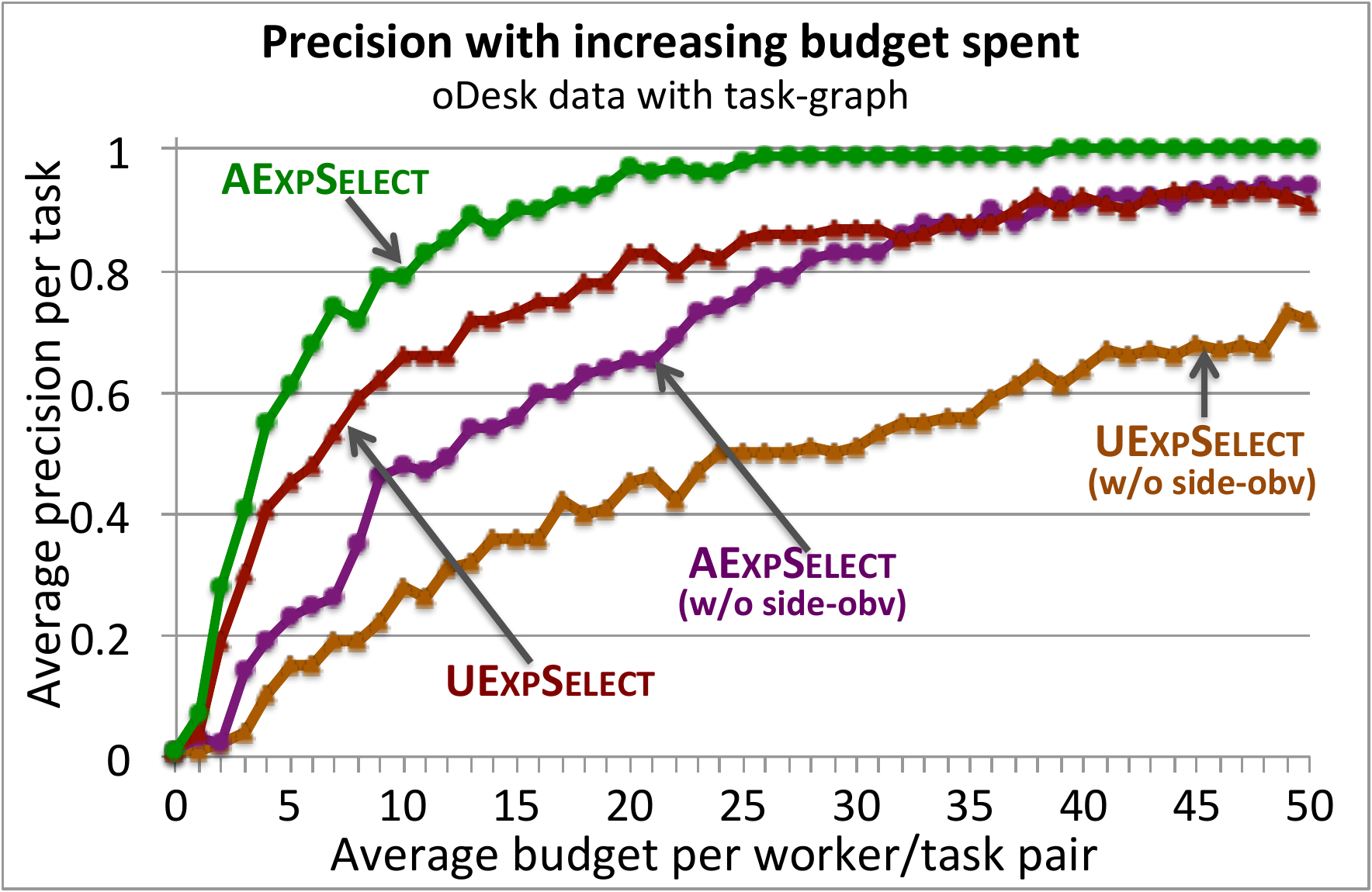}
     \label{fig.odesk2}
   }
   \subfigure[Side-observations over tasks \& workers]{
     \includegraphics[width=0.32\textwidth]{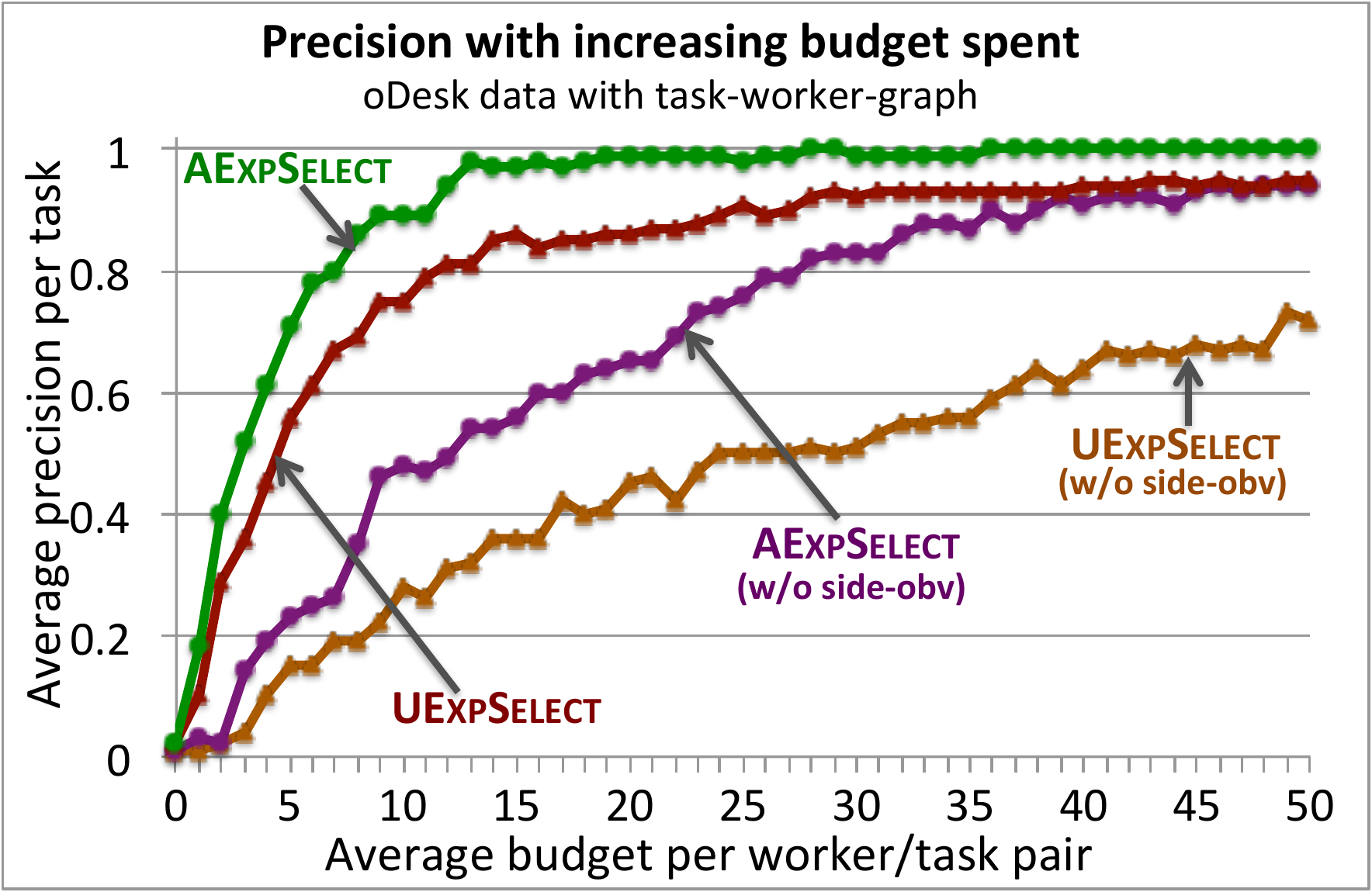}
     \label{fig.odesk3}
   }
\vspace{-1mm}
\caption{Experimental results on oDesk data. In all the plots, budget is varied and metric of average precision is measured. In Figure~\ref{fig.odesk1}, there is no side-observation graph and is equivalent to the plot in Figure~\ref{fig.syn1}. Figure~\ref{fig.odesk2} and Figure~\ref{fig.odesk3} shows the comparison of \aexp and \uexp with their variants without side-observations ($E_o = \emptyset, E_w = \emptyset$ as input).}
\label{fig.resuts1}
\end{figure*}
%%%%%%%%%%%%%%%%%%%%%%%%%%%%%%%%%%%%%%%%%%%%%%%%%%%%%%%%%

%%%%%%%%%%%%%%%%%%%%%%%%%%%%%%%%%%%%%%%%%%%%%%%%%%%%%%%%%
%%%%%%%%%%%%%%%%%%%%%%%%%%%%%%%%%%%%%%%%%%%%%%%%%%%%%%%%%
\subsection{Experimental Setup and Datasets}
We compare the performance of adaptive algorithm \aexp against the uniform exploration based algorithm of \uexp. Furthermore, we quantify the effect of side-observations by comparing these two algorithms with their variants without side-observation graphs (setting $E_o = \emptyset, E_w = \emptyset$ as input).

 %%%%%%%%%%%%%%%%%%%%%%%%%%%%%%%%%%%%%%%%%%%%%%%%%%%%%%%%%
\subsubsection{Metrics and parameters.}
%implementation choices
The primary metric is the quality of the team output by the algorithm for given budget, measured through \emph{i}) average precision, and \emph{ii}) average performance gap, as defined next. For a given output $S$, and any task $o_j$, the precision for task $o_j$ is defined to be 1 if $S$ contains an $\epsilon$-optimal worker for task $o_j$, \emph{i.e.}, $|S \cap S_{(\epsilon, j)}| \geq 1$, else 0. The performance gap for a task $o_j$ is defined to be $(\mu_{(i^*,j)} -  \operatorname*{max}_{w_i \in S} \mu_{(i,j)})$. We report the average precision and average performance gap over all the $M$ tasks for the team output by the algorithm for a given budget.

The primary quantity that we vary in the experiments is the total number of tests performed or budget spent by the algorithm. For ease of interpretation, we shall use the unit of the average budget spent per worker/task pair. Also, we shall report results by varying the hardness of the problem instance (Figure~\ref{fig.syn3}). For a given task $o_j$, we used the notion of hardness given by $\Delta_{j}^{min} = \operatorname*{min}_{i \in [1 \ldots N]} \Delta_{(i,j)}$. We vary average value of the gap $\Delta_{j}^{min}$ over tasks (\emph{i.e.}, $\frac{1}{M} \cdot \sum_{j \in [1 \ldots M]}  \Delta_{j}^{min} $) by creating different datasets and measuring the performance of different algorithms for a fixed budget. 

The PAC parameters $\epsilon$ and $\delta$ are fixed for all of the reported experiments and set to $0.05$. The number of tasks is $M=10$ and total number of workers is $N=200$. In all of the experiments with varying budget, the average $\Delta_{j}^{min}$ over tasks is fixed to $0.25$, with $\Delta_{j}^{min}$ for a task $o_j$ uniformly sampled in the range from $[0.01, 0.5]$. For the experiment in Figure~\ref{fig.syn3} where the average $\Delta_{j}^{min}$ is varied, the average budget per worker/task pair is fixed to $20$, \emph{i.e.}, equivalent to total budget of $M \cdot N \cdot 20$. The values of the performance matrix $\mu$ are scaled to lie in the range $[\mu^{min}, \mu^{max}]$ where $\mu^{min}=0.1$ and $\mu^{max}=0.9$. We assume a Bernoulli feedback model, {\emph i.e.}, for $\mu_{(i,j)}$, assigning task $o_j$ to worker $w_i$ yields a feedback value of $1$ with probability $\mu_{(i,j)}$ and $0$ otherwise. All the results are reported as an average of $10$ iterations of the algorithms.

%%%%%%%%%%%%%%%%%%%%%%%%%%%%%%%%%%%%%%%%%%%%%%%%%%%%%%%%%
\subsubsection{Synthetic data.}
We created synthetic data for $N=200$ workers and $M=10$ tasks as follows. For each task $o_j$, we sampled $\Delta_{j}^{min}$  uniformly at random from range $[0.01,0.5]$ (to have average $\Delta_{j}^{min}=0.25$). Then, to create the performance vector $\mu_{(.,j)}$ for $N$ workers (corresponding to a column in the performance matrix in Figure~\ref{fig.approach}), we sampled $(N-1)$ values in the range $[\mu^{min}, \mu^{max} - \Delta_{j}^{min}]$ and one value (of the best worker) is set to $\mu^{max}$. These $N$ values are then randomly permuted and assigned to the $N$ workers for task $o_j$. This process is repeated for each of the $M$ tasks independently. For the synthetic experiments,  we didn't use side-observations, equivalent to having $E_w = \emptyset$ and $E_o = \emptyset$. For the experiment reported in Figure~\ref{fig.syn3}, we created 4 more variants of the synthetic data by varying average $\Delta_{j}^{min}$ as $[0.25, 0.20, 0.15, 0.10, 0.05]$.

%%%%%%%%%%%%%%%%%%%%%%%%%%%%%%%%%%%%%%%%%%%%%%%%%%%%%%%%%
\subsubsection{oDesk data.}
The primary purpose of using data from oDesk is to be able to obtain real-world distributions of the performance matrix, as well as a realistic way of creating the side-observation graphs. oDesk has over $2.7$ million freelancers and $0.5$ million job requesters worldwide. We used the publically available API\footnote{\url{https://developers.odesk.com/}} from oDesk to obtain the data below. In oDesk platform, each posted task or job is assigned to a predefined taxonomy by the job requester. There are $12$ top-level categories of the tasks and about $90$ second-level categories. We took $M=10$ tasks, with 4  tasks in the top-level category \emph{Design \& Creative}, 3 tasks in the top-level  category \emph{Translation} and $3$ tasks in \emph{Data Science \& Analytics}. We note that this choice is arbitrary, and does not effect the reported results qualitatively. We also performed experiments on other variants of the oDesk datasets that considered different sets of task types.
%%%, though do not report the results for other datasets here.

Each worker in the oDesk has a profile with rich metadata available via an API. In particular, the fields that are of particular interest to us include: \emph {i}) the ``skills" (a set of free-form text tags that workers can assign to themselves); \emph {ii}) feedback score based on previous tasks completed; \emph {iii}) number of hours worked; and \emph {iv}) the top-level categories of the tasks completed by the workers, based on which the feedback score is aggregated. We crawled a sample of 200 workers by issuing a specific query \footnote{\texttt{\{`hours': `[100 TO 10000]', `skills': `cartooning OR machine-learning OR translation'\}}}. 
%%%\footnote{\texttt{\{`hours': `[100 TO 10000]', `skills': `cartooning OR translation-english-chinese OR machine-learning'\}}}. 
The skills in this query were chosen so as to ensure that the completed jobs by the workers in the retrieved list possibly have some overlap with the top-level categories of the $M$ tasks, otherwise, this overlap would be low for a randomly retrieved list of workers. In realistic setting, this overlap is expected as workers bid for tasks based on their skills and job profile. The number of hours worked was set to a minimum of $100$ to ensure there is sufficient feedback available for the workers, given that the feedbacks are generally sparse.

We created the side-observation graphs as follows. We add an edge between two tasks $o_x$ and $o_z$, \emph{i.e.},  $\{o_x, o_z\} \in E_o$,  if these two tasks belong to the same top-level category. In our setting, this would result in $3$ disconnected cliques among the $10$ tasks. For the workers, we computed the Jaccard's coefficient between the skills of any two workers. We add an edge between two workers $w_x$ and $w_z$, \emph{i.e.},  $\{w_x, w_z\} \in E_w$,  if the Jaccard's coefficient between $w_x$ and $w_z$ is above a certain threshold (chosen to be $0.3$ for the reported results). Next, we create the performance matrix from the feedback scores in a similar manner to the approach we took with the synthetic data. First, for each task $o_j$, we sampled $\Delta_{j}^{min}$ uniformly at random from range $[0.01,0.5]$. Then, for a given worker $w_i$ and task $o_j$, we look at the feedback score of $w_i$ obtained in the historically completed tasks which belong to top-level category same as that of $o_j$. Note that this feedback score is in the range of $[0, 5]$ rating. When available, this feedback score is used for $\mu_{(i,j)}$, else feedback score is randomly sampled from $[0,3]$. These feedbacks are then scaled to lie in the range $[\mu^{min}, \mu^{max} - \Delta_{j}^{min}]$, except for best worker for $o_j$, whose $\mu_{(i,j)}$ is set to $\mu^{max}$. This process is repeated for each of the $M$ tasks independently.

%%%from the list of workers retrieved 
%%
%%%%%\texttt{\{`skills': `cartooning OR translation-english-chinese OR machine-learning', `hours': `[100 TO 10000]'\}} 
%%
%%%(otherwise, this overlap would be low given the large number of categories)
%%%otherwise the feedback would surely be zero given large number of workers who have only completed very few jobs and didn't receive any feedback.
%%%%%%%%%%%%%%%%%%%%%%%%%%%%%%%%%%%%%%%%%%%%%%%%%%%%%%%%%
%%%%%%%%%%%%%%%%%%%%%%%%%%%%%%%%%%%%%%%%%%%%%%%%%%%%%%%%%
\subsection{Results}
We now discuss the findings from our experiments.

{\bf Varying budget and measuring precision.}
Figure~\ref{fig.syn1} and Figure~\ref{fig.odesk1} shows the results for varying the average budget spent per worker/task pair, and how it leads to increased precision of the team selected by \uexp and \aexp. For these results, the average  $\Delta_{j}^{min}  = 0.25$, and is same for both the synthetic data (Figure~\ref{fig.syn1}) and the oDesk data (Figure~\ref{fig.odesk1}). For both the datasets, \aexp shows significantly faster convergence towards selecting the optimal team. For instance, in Figure~\ref{fig.syn1}, \aexp achieved over $90\%$ precision (getting the $\epsilon$-optinal worker for $9$ out of $10$ tasks) at budget of $20 \cdot M \cdot N$, whereas \uexp requires substantially much more budget to achieve same precision. The difference in performance of \aexp or \uexp across synthetic and oDesk datasets is simply attributed to the different distribution of the workers' performances across the datasets. In particular, in the oDesk data, the performance values of the workers are more skewed towards higher values making it more challenging problem instance, in comparison to the synthetic data where the performance values are sampled uniformly.

{\bf Varying budget and measuring performance gap.}
Figure~\ref{fig.syn2} shows an alternate view of the corresponding result in Figure~\ref{fig.syn1}. While Figure~\ref{fig.syn1}  reported $0/1$ loss, Figure~\ref{fig.syn2} reports the average of the actual performance gap of the best worker for a task in the output set compared to the best worker in full set.

{\bf Varying hardness of problem instance.}
In Figure~\ref{fig.syn3}, budget is kept fixed to $20$ per worker/task pair, and average $\Delta_{j}^{min}$, quantifying the hardness of the problem instance, is changed from $0.25$ to $0.05$. The gain of adaptive assignments in \aexp compared to \uexp is consistent, though both the algorithms degrade in performance as expected.

{\bf Effect of exploiting side-observations.}
In Figure~\ref{fig.odesk1}, there is no side-observation graph, ($E_o = \emptyset, E_w = \emptyset$), and the $\widetilde{ \gamma}_{G_{wo}}$ is simply equal to $M \cdot N$. In Figure~\ref{fig.odesk2}, there is side-observation graph over the tasks as described in the  data generation, however no graph is used over workers ($E_w = \emptyset$). The $\widetilde{ \gamma}_{G_{wo}}$ is this case as computed by greedy algorithm is equal to $600$. Figure~\ref{fig.odesk3}  shows results which considers side-observation graphs over both tasks and workers, with  $\widetilde{ \gamma}_{G_{wo}} = 386$.  Both the algorithms see a significant boost in terms of faster learning by exploiting the side-observations. Furthermore, we can see that the boost in performance by adding side-observations is more for \uexp compared to \aexp, as discussed during the theoretical performance analysis of the algorithms.
 
%%%%%%%%%%%%%%%%%%%%%%%%%%%%%%%%%%%%%%%%%%%%%%%%%%%%%%%%%
%%%%%%%%%%%%%%%%%%%%%%%%%%%%%%%%%%%%%%%%%%%%%%%%%%%%%%%%%

%% file: conclusion.tex
%%%%%%%%%%%%%%%%%%%%%%%%%%%%%%%%%%%%%%%%%%%%%%%%%%%%%%%%%
%%%%%%%%%%%%%%%%%%%%%%%%%%%%%%%%%%%%%%%%%%%%%%%%%%%%%%%%%
\section{Conclusions and Future Work}\label{sec.conclusions}
We presented an algorithmic approach to tackle the challenge of the efficient hiring of teams of workers, as faced by recruiters for contract-based crowdsourcing. By casting these budgeted decision-theoretic problems as an instance of online learning for best action selection, we designed algorithms with PAC bounds, and further extended them to exploit the commonalities among the tasks and the workers. Our methodology and results present an interesting direction of continued research for the problem of hiring a team for contract-based crowdsourcing.  

We see several interesting directions in which the current work can be extended.  In particular, we used a simple notion of quantifying the optimality of the team. We see promise in extending the results to incorporate more complex relations among team members, such as the matching of task types within teams to balance the workload, capturing diminishing returns of growing teams, learning and representing  costs associated with communication and coordination among people with different skills and abilities (including collaborative competency), and other combinatorial constraints, as an interesting direction for future work. Furthermore, we are interested in developing more realistic models of side-observations and performing real-world experiments using those models.
%%%%%%%%%%%%%%%%%%%%%%%%%%%%%%%%%%%%%%%%%%%%%%%%%%%%%%%%%
%%%%%%%%%%%%%%%%%%%%%%%%%%%%%%%%%%%%%%%%%%%%%%%%%%%%%%%%%